# Connectional-Style-Guided Contextual Representation Learning for Brain Disease Diagnosis

Gongshu Wang, Ning Jiang, Yunxiao Ma, Tiantian Liu, Duanduan Chen, Jinglong Wu, Guoqi Li, *member*, *IEEE,* Dong Liang, *senior member*, *IEEE*, and Tianyi Yan, *member*, *IEEE*

*Abstract*—Structural magnetic resonance imaging (sMRI) has shown great clinical value and has been widely used in deep learning (DL) based computer-aided brain disease diagnosis. Previous approaches focused on local shapes and textures in sMRI that may be significant only within a particular domain. The learned representations are likely to contain spurious information and have a poor generalization ability in other diseases and datasets. To facilitate capturing meaningful and robust features, it is necessary to first comprehensively understand the intrinsic pattern of the brain that is not restricted within a single data/task domain. Considering that the brain is a complex connectome of interlinked neurons, the connectional properties in the brain have strong biological significance, which is shared across multiple domains and covers most pathological information. In this work, we propose a connectional style contextual representation learning model (CS-CRL) to capture the intrinsic pattern of the brain, used for multiple brain disease diagnosis. Specifically, it has a vision transformer (ViT) encoder and leverages mask reconstruction as the proxy task and Gram matrices to guide the representation of connectional information. It facilitates the capture of global context and the aggregation of features with biological plausibility. The results indicate that CS-CRL achieves superior accuracy in multiple brain disease diagnosis tasks across six datasets and three diseases and outperforms state-of-the-art models. Furthermore, we demonstrate that CS-CRL captures more brain-network-like properties, better aggregates features, is easier to optimize and is more robust to noise, which explains its superiority in theory. Our source code is available on: https://github.com/NingJiang-git/CS-CRL

*Index Terms*—Transformers, MRI, self-supervised, brain diseases diagnosis.

## I. INTRODUCTION

AFTER their initial success in computer vision, deep learning (DL) models have rapidly gained traction in computer-aided diagnosis (CAD). Structural magnetic resonance imaging (sMRI) can noninvasively describe brain volume, density, and morphology, from which various CAD approaches have been proposed. Most sMRI-based DL methods leverage labeled data with limited samples directly to train the model and use a convolutional neural network (CNN) backbone with compact filters to perform the same processing on every partition of an image. These methods have the following two limitations:

(1) Supervised representation learning can quickly acquire informative representations related to labels but may neglect other important information, leading to poor understanding of the brain that suffers from generalization error, spurious correlations, and adversarial attacks[1].

(2) The strong inductive bias in CNNs effectively leverages parameters to facilitate training but has disadvantages in capturing global structures[2, 3]. In addition, compared to natural images, sMRIs are of low contrast and are easily corrupted by noise or artifacts[4]. However, CNNs are sensitive to high frequency signals, which are weakly robust against nuisance factors[3, 5].

Therefore, sMRI-based end-to-end CNN models have yielded satisfactory accuracy only in the diagnosis of neurological diseases, such as Alzheimer's disease (AD) and mild cognitive impairment (MCI)[6-9], where brain atrophy and local pathological features are significant, but they have not seen a breakthrough in the diagnosis of psychological diseases with subtle and discrete atrophy, such as schizophrenia (SZ) and attention deficit hyperactivity disorder (ADHD), which require further extraction of task-specific features from sMRI [10].

To pursue a high-performance model, it is necessary to

This work was supported part by National Natural Science Foundation of China (U20A20191, 61727807, 82071912, 12104049); the Beijing Municipal Science & Technology Commission (Z201100007720009); the Fundamental Research Funds for the Central Universities (2021CX11011); and the China Postdoctoral Science Foundation (2020TQ0040) *(Co-first authors: Gongshu Wang and Ning Jiang; Corresponding author: Tianyi Yan).*

G. Wang, N. Jiang, Y. Ma, T. Liu, D. Chen, J. Wu and T. Yan are with School of Life Sciences, Beijing Institute of Technology, 100081 Beijing, China (e-mail: gongshu@bit.edu.cn; ningjiang@bit.edu.cn; mayunxiao@bit.edu.cn; tiantian2bit@bit.edu.cn; duanduan@bit.edu.cn; wujl@bit.edu.cn; yantianyi@bit.edu.cn ).
G. Li is with Institute of Automation, Chinese Academy of Sciences, Beijing, China (E-mail: guoqi.li@ia.ac.cn)
D. Liang is with Research Center for Medical AI, Shenzhen Institutes of Advanced Technology, Chinese Academy of Sciences, Shenzhen, China (e-mail: dong.liang@siat.ac.cn).

comprehensively understand the intrinsic pattern with biological significance of the brain[11], because they are robust across domains and cover semantic information relevant to most diseases. The brain is known to be a strong connectome. Cognitive, emotional and motor functions are realized through interactions between neurons[12]. We know that one brain region contains a lot of biological information about other regions. The abnormalities in one region may have a ripple effect through others, and the distribution of pathology involves context regarding the position and structure of lesions about healthy tissue, so inter-regional dependencies represent more physiological significance than local features. Previous studies have established gray matter (GM) brain networks and demonstrated that it is a reliable and robust morphological feature of the brain structure[13, 14] and is valuable for the diagnosis of brain diseases[15, 16]. These studies show that sMRI can describe the potential connections between regions, which to some extent reflects the interaction between neurons. Therefore, using sMRI to diagnose psychological diseases instead of functional magnetic resonance imaging (fMRI) is quite feasible and of great significance. In this study, we pursue a comprehensive and advanced brain representation learning method to extract the global context from sMRI to capture the critical connectional properties in the brain.

Recently, [17] proposed vision transformer (ViT) by formulating image classification as a sequence prediction task, thereby capturing long-range dependencies within the input image leveraging multi-head self-attentions (MSAs)[18]. The MSAs have proven to be sensitive to the global context, which is beneficial for capturing interactions between regions with higher-order physiological significance. In addition, ViTs are robust against data corruptions, image occlusions[19] and high-frequency noises[3, 20]. Theoretically, the ViT is highly matched with our pursuit in brain sMRI analysis.

In this study, we aim to exploit the potential of ViTs for deep representations of the brain and, accurately capture the crucial connectional properties as biomarkers for diagnosis. Therefore, a challenging pretext task, which leverages a small fraction of sMRI to infer the entire image, enables the ViT to effectively integrate context and make a deep understanding of brain pattern. Masking image modeling (MIM) is an effective visual representation learning strategy that can meet our requirement. The basic idea of MIM is masking and reconstructing: masking a set of image patches and reconstructing them at the output. MIM forces the network to infer the masked target by aggregating visible context. Recent studies have demonstrated that MIM, a simple self-supervised learning (SSL) method, performs superiorly on a variety of tasks[21-23].

As for a model that adapts to the characteristics of brain sMRI data and can extract biological biomarkers accurately, some biological prior knowledge is needed. In general, there are strong interactions between some paired brain regions and weak interactions between others. Similar basic properties exist in all regions, but each brain region has a unique composition and function. Therefore, this simple but critical connectional property of the brain can be used as a regularization, making it easy for the model to learn the representations as we expect.

Motivated by the abovementioned analysis, we propose a brain representation method for multiple brain disease diagnosis, named connectional style contextual representation learning model (CS-CRL). It is based on the ViT, trained by a tailor-made MIM proxy task and incorporates a specific brain connectional prior. First, the sMRI is cut into 3D patches and a fraction of patches are fed into a ViT encoder. Second, two projectors are used to distill strong connection-related and weak connection-related semantic information and establish latent representations from the encoded patches. Specifically, the Gram matrices are used to guide the learning objectives of the projectors. Then, the two sets of latent representations are sent to a shared ViT decoder. Finally, the difference of the two output images is identified as the reconstructed image. Our method is used to pre-train the model and enable the encoder to capture biologically meaningful context in the brain that is invariant to domains. When transferred to downstream tasks, these strong representations can be easily translated into high-order disease-specific biomarkers with only simple fine-tuning. Our proposed CS-CRL achieved superior accuracy, generalization and explanation in brain disease diagnosis.

Our main contributions are summarized as follows:

1) We point to the potential of ViTs in brain sMRI analysis and propose CS-CRL that integrates computational and biological principles into a framework. It leverages a specific MIM task and a crucial brain connectional regularization to guide the ViT for representation learning and is therefore more adaptive for capturing connectional patterns in the brain.

2) We used the masked autoencoder (MAE), a state-of-the-art MIM method, as a strong baseline to explore the theoretical superiority of our CS-CRL and demonstrated that CS-CRL is easier to optimize and fine-tune, better establishes biologically plausible interregional dependencies, removes more high-frequency noise and aggregates more stable representations.

3) We verify the generalization and transferability of our CS-CRL by fine-tuning the pre-trained model for diagnosis across three diseases and three transfer learning tasks. It shows significantly better performance compared with a ViT trained from scratch, a ViT pre-trained by MAE and state-of-the-art CNNs. We revealed that a simple strategy: incorporating suitable biological priors as the constraint, is more effective than complicating the architecture of the model.

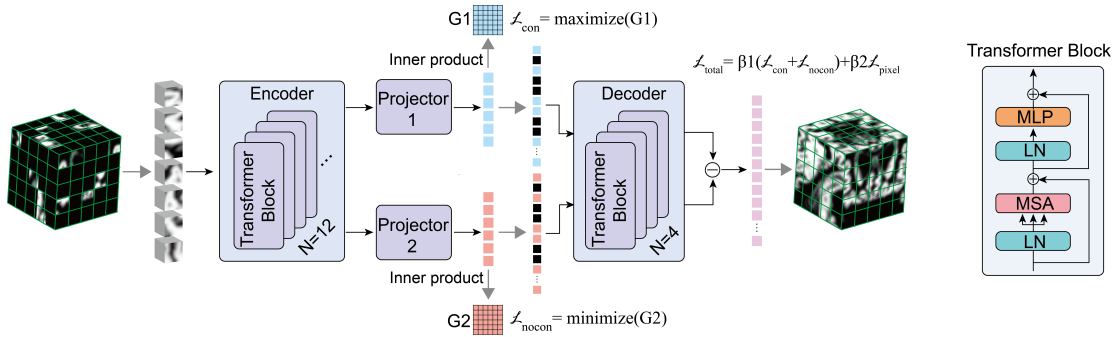

**Fig. 1.** Architecture of our CS-CRL. sMRI is cut into 3D patches. A large random subset of image patches (e.g., 76%) is masked out. The encoder is applied to the small subset of visible patches. G1 and G2 are Gram matrices of two latent spaces. One is used to guide the connectome semantic representation in the latent space, and the other is used to guide the nonconnectome semantic representation in the latent space. Mask tokens are introduced after the encoder. A decoder is used to process two sets of latent representations (encoded patches and mask tokens). The original sMRI results from differencing the two outputs.

## II. RELATED WORK

### A. sMRI-based CAD for Brain Diseases

CAD solutions assist specialists in effectively evaluating large-scale medical data, improving diagnosis accuracy and consistency with less analysis time. Traditional machine learning (ML) and DL have been widely used to develop CADs.

Traditional ML methods rely heavily on feature extraction and selection. Usually, studies extract brain volume, cortical thickness and surface area, regional gray matter and white matter features from sMRI and use PCA, Fisher, cluster, etc., to select features[24-29]. However, selecting proper methods for feature extraction and selection requires extensive knowledge of feature engineering and artificial intelligence (AI), and the greater the complexity of data, the more difficult it is to select the optimal features[30, 31]. In addition, feature extraction and selection depend on certain subjectivity, which may hinder the comparison of results across studies[32].

DL combines feature extraction and selection steps into an automatic feature extraction step and decouples from additional domain-specific knowledge. For DL-based CAD, the main challenge is how to design suitable models to automatically extract disease biomarkers for diagnosis. Recently, DL-based models have been widely used for brain neurological and psychological diseases diagnosis. In [6], a hierarchical full CNN was proposed for AD diagnosis, which could learn multi-scale feature representations. In [33], an attention-guided deep-learning framework was proposed to extract multilevel discriminative sMRI features for dementia diagnosis. In [8], deep multi-instance learning (MIL) models were constructed using local image patches pre-identified by anatomical landmarks for AD classification and MCI conversion prediction. In [7], a dual attention MIL model was proposed for AD diagnosis, which can capture local and global structural features from sMRI. In [34], 3D CNN models with sequential architecture, an inception module and a residual module were used for SZ diagnosis. In [35], a dilated 3D CNN method was used for recognizing patterns across long-distance brain areas in ADHD sMRI. All of these methods are based on CNNs and aim to accurately identify pathological locations but ignore abnormal long-term connection between regions, so these models only superficially understand the brain. Moreover, unlike neurological diseases with obvious alterations, brain atrophy is subtle and discrete in psychiatric diseases such as SZ and ADHD. Therefore, it is a challenge to design efficient DL-based CAD models for automatically extracting pathological features in psychiatric diseases.

### B. ViTs in Medical Imaging

ViTs have recently shown their superiority in natural image processing, and [36] demonstrated their enormous potential in medical imaging, such as medical segmentation, classification, detection and synthesis.

In medical segmentation, many studies (i.e., Trans-UNet[37]) and LeViT-UNet[38]) have adopted architectures combining U-Net and transformers because of the good performance and widespread application of U-Net in medical segmentation. In contrast, [39] employed the swin-transformer[40] block to replace the convolution block in U-net and demonstrated that a pure transformer-based architecture outperforms CNN-based methods or a combination of convolution and transformer.

Transformers are also introduced into medical synthesis to build a more advanced generative adversarial network. In [41], BERT[42] was trained to predict real vs. generated PET and to predict masked values (MLM) from the max-pooled images.[43] (VTGAN) adopted a ViT-based generative adversarial network to synthesize FA images and predict retinal degeneration.

In medical classification, the pure ViT is rarely used, especially for sMRI analysis, because of its poor trainability and unsatisfactory performance. Even worse, most recent ViT-based models used 2D images such as breast ultrasound images[44] and chest X-rays[45] or 2D slices of 3D volumes[46, 47] as their input. However, such a practice catastrophically damages the cross-plane contextual information in the 3D volume, especially the critical anatomical information. In contrast, we adopt the 3D volume of brain sMRI as the input and design a self-supervised pre-training model to overcome the limitations of previous models.

## C. Pre-training in Medical Imaging

Due to the lack of labeled medical images, many studies pre-train their models on large-scale natural image datasets (such as ImageNet) and transfer the pre-trained models to medical imaging tasks[48] [49]. Transfer learning methods have also been used to pre-train ViT models, such as [50]. However, because of the domain gap between natural and medical images, these studies cannot always achieve their desired performance.

Self-supervised learning (SSL) is considered to be of great potential in medical imaging, showing superior performance compared with conventional transfer learning methods. [51] pre-trained and fine-tuned their model in the same data domain with a proxy task. Then, the model was fine-tuned on a supervised segmentation task and performed well. [52] pre-trained the network with DINO[53] and showed that SSL works better than transfer learning and supervised learning. Regrettably, most self-supervised ViT-based models expect 2D input and fall into the trap of losing the cross-plane contextual information.

Of course, there are studies for self-supervised pre-training using 3D medical images. [54] designed five proxy tasks for pre-training on unlabeled 3D medical images. Their method was evaluated on three downstream tasks and performed competitively compared with state-of-the-art methods with less computation. [55] proposed a generic autodidactic model and demonstrated that their method outperforms any 2D approaches, both models transferred from ImageNet and the 2D version of their model. However, these methods are designed for CNN-based models, which are not exactly appropriate for ViTs. More importantly, due to the heterogeneity of natural and medical images, these studies, which transfer the self-supervised pre-training methods from general computer vision tasks directly, are limited by a lack of biological prior knowledge and cannot achieve satisfactory performance. In fact, there have been few articles about pre-training ViTs for brain image analysis until now, especially for classification tasks. Therefore, we propose a self-supervised ViT-based model, which consider the connectional information in the brain and is pre-trained by a specific MIM task, to avoid the limitations mentioned above.

## III. METHODS

### A. CS-CRL

Here, we propose a brain representation learning model, which is based on ViT, combines MIM and integrates the connectional properties of the brain as a prior. The main goal of our model is to train a strong encoder that can be transferred to multiple brain disease diagnosis and transfer learning tasks. Therefore, our model uses a ViT as the backbone to focus on global context information in the brain, a challenging MIM proxy task to force the encoder to make a deep and comprehensive understanding of sMRI, and imposes biological priors to enable it to acquire significant biological representations.

The encoder consists of a patch block and a series of transformer blocks, and the decoder also consists of a series of transformer blocks. Similar to some advanced MIMs, we use an encoder-decoder asymmetric structure, in which the decoder is lightweight, because our focus is on encoder performance. However, unlike classical autoencoders, our CS-CRL leverages two parallel projection layers to distill distinct styles of latent representations, and a shared decoder maps latent representations separated by style into two distinct components of the original image. Specifically, with regard to describing the style of the latent representations, we use latent representations from two projection layers to build Gram matrices, which can be considered descriptions of the semantic information of the input image[56]. Furthermore, to regularize the model with the connectional properties, we additionally design a semantic loss function, which makes the mean value of one Gram matrix infinitely increase and the mean value of the other Gram matrix infinitely decrease. As a result, one matrix focuses on regions that are interconnected and extract semantically similar components from them. The other matrix focuses on regions that are relatively independent and extract semantically dissimilar components from them.

Considering the infinite potential and weak inductive bias of the ViT, designing challenging proxy tasks and imposing the guidance that are biologically meaningful and adapted to MSA can significantly alter the mode of establishment of inter-regional dependencies and aggregation of features. Thus, the encoded sMRI is not a data-driven low-level image feature, but a holistic understanding of the biological properties of the brain. When the encoder is transferred to downstream tasks, it can detect biomarkers more effectively and efficiently.

See detailed composition of CS-CRL in this section.

#### 1) ViT

We use a ViT[17] as the backbone for both the encoder and decoder. A ViT is composed of a patch embedding layer, position embedding and a number of transformer blocks.

Patch embedding: For a 3D volume $x \in \mathbb{R}^{H \times W \times D \times C}$, we first reshape it into a sequence of flattened 3D patches with a fixed size $(P, P, P)$ without overlapping, namely, $x_p \in \mathbb{R}^{N \times (P \cdot P \cdot P \cdot C)}$, where $N = H \times W \times D/P^3$ is the total number of patches, and $C$ is the input channel. Then, a linear projection layer with learnable parameters is employed to map them into patch embeddings. This step is implemented using 3D convolutions with kernel size $(P, P, P)$ and stride $P$. We use $(10,10,10)$ as the fixed patch size and project these patches into embeddings with 1000 as the hidden size.

Position embedding: To retain positional information for embedded patches, the vanilla ViT adopts 1D learnable embeddings that are added to the patch embeddings. However, we experimentally find that the 1D learnable position embeddings perform equally as the sin-cosine version of position embeddings but increase the number of parameters and computations. Therefore, we use sin-cosine position encoding in the pre-training and fine-tuning stages.

Transformer Block: A transformer block[18] consists of alternating layers of multi-head self-Attention (MSA) [18] and multi-layer perceptron (MLP) blocks. An MLP is composed of two FC layers with a GELU non-linearity, and a dropout layer at the end. The main architecture can be described as follows:

$$z'_l = MSA(LN(z_{l-1})) + z_{l-1} \tquad (1)$$

$$z_l = MLP(LN(z'_l)) + z'_l \tag{2}$$

where $LN$ is LayerNorm and $z_l$ denotes the feature map of the $l_{th}$ transformer block.

Note that we use global average pooling for the output of the transformer blocks as the basis of classification, which is the input to the classification head consisting of a single linear layer.

**2) Masking**

The representation of all flattened patches of an image is referred to as a set of tokens, which can be formulated as $v_{all} \in \mathbb{R}^{N \times L}$, where $L = P^3 C$ is the length of 1D embeddings. For masking, we choose a high mask ratio $m$ (76%) and divide $v_{all}$ into two parts, namely, $v_{vis} \in \mathbb{R}^{N_1 \times L}$ and $v_{mask} \in \mathbb{R}^{N_2 \times L}$, where $N_1 = N \times (1 - m)$, $N_2 = N \times m$. Only $v_{vis}$ will be used as the input to the encoder in pre-training.

**3) Encoder**

The encoder of our model is essentially a ViT composed of 12 transformer blocks using 12 heads for the MSA block. The encoder process can be described as:

$$z_{vis} = e + v'_{vis} \tag{3}$$
$$z'_{vis} = Blocks(z_{vis}) \tag{4}$$

where $v'_{vis}$ is mapped from $v_{vis}$ via a linear projection layer in the patch embedding block, $e$ is the position embedding, and $z'_{vis}$ is the output of the transformer blocks, which are considered to be the latent representations of the input image.

**4) Encoder to Decoder**

In CS-CRL, two different, learned linear projection layers are used to map the encoded patches of our encoder into embeddings in latent space, which can be described as:

$$z''_{vis_i} = project_i(z'_{vis}) \tag{5}$$

where $i = 1,2$. We expect these two latent spaces to represent connectome and nonconnectome semantic information.

We compute the Gram matrices $G_{vis_i} \in \mathbb{R}^{N_1 \times N_1}$ of $z''_{vis_i}$:

$$G_{vis_i} = z''_{vis_i} \cdot z''^T_{vis_i} \tag{6}$$

A Gram matrix computes the inner product between vectors or embeddings, where each element describes the similarity between two encoded patches. A Gram matrix describes the overall style of images and is widely used in neural style transfer[56, 57]. Graph autoencoder (GAE) also uses a similar method to represent the connection probability between nodes[58]. In our work, we use Gram matrices to describe style information in latent space.

We believe that all features (nodes) in the strong connectome are highly correlated with each other, while all features (nodes) in the weak connectome are not. Thus, the criteria on the two latent spaces are as follows:

$$z''_{vis_1} = argmax(\sigma(G_{vis_1})) \tag{7}$$
$$z''_{vis_2} = argmin(\sigma(G_{vis_2})) \tag{8}$$

where $\sigma(\cdot)$ is the sigmoid activation function. This process can also be seen as style transfer. In the optimization process, there is a trade-off between the two different styles to enable the model to be in dynamic equilibrium, which is conductive to the extraction and separation of various semantic information.

**5) Decoder**

Different from the encoder, the decoder is fed with all of the tokens, including $z''_{vis_1}, z''_{vis_2}$, the representations of $v_{vis}$ from the encoder and the masked tokens $z_{mask}$, which are learnable tokens put in the positions of $v_{mask}$. We implement the decoder with a ViT containing 4 transformer blocks using 6 heads in the MSA block. The decoder process can be described as:

$$z_{all_i} = z''_{vis_i} \oplus z_{mask} \tag{9}$$
$$z'_{all_i} = e' + z_{all_i} \tag{10}$$
$$y_{all_i} = Blocks(z'_{all_i}) \tag{11}$$
$$(y_{vis}, y_{mask}) = y_{all} = y_{all_1} - y_{all_2} \tag{12}$$

where $e'$ is the position embedding for all of the tokens and $y_{all_i}$ are the outputs of transformer blocks. $y_{all}$ is the output of our decoder, which can be divided into $y_{vis}$ and $y_{mask}$.

Note that our decoder is only used in pre-training to assist the encoder in growing into a connectional semantic information extractor we desire.

**6) Loss Function**

The task in the pre-training stage is to predict the original pixels in the masked patches and regularize the latent space with biological prior. First, there is pixel-level loss, which is essentially calculated as the mean square error (MSE).

$$L_{pixel} = \| y_{mask} - v_{mask} \| \tag{13}$$

The learning objectives for the connectome and nonconnectome semantic information can be formulated as:

$$L_c = -log(\sigma(G_{vis_1})) \tag{14}$$
$$L_{nc} = -log(1 - \sigma(G_{vis_2})) \tag{15}$$

Therefore, the overall objective function is:

$$L_{all} = \beta_1 L_{pixel} + \beta_2 (L_c + L_{nc}) \tag{16}$$

where $\beta_1$ and $\beta_2$ are commitment loss hyperparameters set to 0.99 and 0.005. There is a trade-off between pixel loss and semantic loss, which enables the latent space to not only be used to accurately infer the masked pixels but also semantically interpret the biological nature of the brain.

TABLE I
DEMOGRAPHIC DETAILS OF THE STUDIED SUBJECTS

| Dataset | Group Type | Gender (Male/Female) | Age (Mean±Std) |
|---|---|---|---|
| IXI | NC | 240/309 | 48.64±16.42 |
| ADNI | AD | 184/157 | 75.26±8.03 |
|  | NC | 182/237 | 73.61±6.13 |
| OASIS-3 | AD | 53/117 | - |
|  | NC | 72/121 |  |
| MCIC | SZ | 78/26 | 34.85±11.12 |
|  | NC | 64/28 | 33.39±12.12 |
| NUSDAST | SZ | 72/24 | 32.84±12.46 |
|  | NC | 53/43 | 31.79±13.45 |
| NYU | ADHD | 103/36 | - |
|  | NC | 51/50 |  |
| KKI | ADHD | 10/15 | 10.04±1.55 |
|  | NC | 13/17 | 9.98±1.29 |

*B. sMRI-datasets*

**1) Datasets**

T1-weighted images in seven public datasets were used in this study (Table. 1), including one dataset for pre-training and other six datasets across three diseases (i.e., AD, SZ, ADHD) for downstream classification tasks.

Healthy subjects in IXI (https://brain-development.org/ixi-

dataset/) were used to construct our pre-training dataset. ADNI [59] (mixed ADNI 1, 2 and 3, https://ida.loni.usc.edu/login.jsp) and OASIS-3 [60] (https://www.oasis-brains.org/) were used for AD diagnostic tasks. MCIC [61] and NUSDAST [62] (http://www.schizconnect.org/) were used for SZ diagnostic tasks. NYU and KKI, the subsets of ADHD-200 [63] (http://fcon_1000.projects.nitrc.org/indi/adhd200/), were used for ADHD diagnostic tasks.

### 2) Image Preprocessing

First, all T1-weighted images were skull-stripped. Second, brain-extracted T1-weighted images were corrected for intensity nonuniformity. Then, voxel-based morphometry (VBM)[64] was used to segment brain tissues, acquiring gray matter (GM). All GM density maps were linearly registered to the standard Montreal Neurological Institute (MNI) GM tissue probability template with a 1.5×1.5×1.5 mm$^3$ voxel resolution. These procedures were performed with FSL package [65]. The human subcortical refers to a deep GM structure, consists of hundreds of unique, small gray matter nuclei[66]. In subcortical areas, neural elements are densely interconnected, and local computations are highly segregated[67, 68]. According to the subcortical mask, we captured a region of 50×50×50 voxels near the center of the image, which almost covered the whole subcortical GM, as features of our study.

Data were excluded if a T1-weighted image failed to complete the tissue segmentation procedure or if tissue segmentation was inaccurate.

TABLE II
HYPERPARAMETERS AND OPTIMIZATION STRATEGY OF CS-CRL

| Procedure | Pretrain. | Finetune. |
|---|---|---|
| Batch size | 8 | 16 |
| Optimizer | AdamW | AdamW |
| LR | × | × |
| Base LR | 1.5E-04 | 1.0E-03 |
| LR decay | cosine | cosine |
| Layer-wise decay | × | 0.75 |
| Weight decay | 0.05 | 0.05 |
| Warmup epochs | 40 | 5 |
| Label smoothing | × | 0.1 |
| Dropout | × | 0.1 |
| Gradient Clip. | × | × |
| Mixup alpha | × | 0.8 |
| Init scale | × | 0.001 |

### 3) Implementation

We performed the pre-raining and fine-tuning stages on four NVIDIA Tesla P100 GPUs following Table 2. In pre-training, we choose samples from the IXI dataset as the training set. Although the reconstruction quality and the total loss can reflect if our network converges at its best to some extent, we still pre-trained the models for 1000, 2000, and 3000 epochs to analyze the effect of pre-training epochs for classification performance. For fine-tuning, a new classifier with randomly reinitialized weights is added to the encoder of the CS-CRL. In within-domain classification, we divide the samples from every dataset into training, validation and test set randomly in a 6:2:2 ratio. In cross-domain classification, we divide the samples from the source domain into training and validation sets randomly in a 4:1 ratio, and all of the samples from the target domain are used as the test set.

### 4) Competing Methods

We evaluate the performance of our proposed CS-CRL on the five above-mentioned datasets compared with five methods, i.e., a conventional baseline for patch-based models (i.e., PLM), a state-of-the-art sMRI classification method used for AD and MCI diagnosis (i.e., DA-MIDL), a classical convolutional neural network (i.e., ResNet-50), a ViT trained from scratch and a ViT pre-trained by MAE.

Patch level method (PLM): sMRI data are first divided into non-overlapping patches, and then the average gray matter density of each patch is calculated as the input feature of the model. Finally, a linear SVM is used to construct the classifiers.

Dual attention multi-instance deep learning (DA-MIDL) model: The DA-MIDL model consists of Patch-Net with a spatial attention block, attention multi-instance learning pooling and an attention-aware global classifier[7].

ResNet-50: ResNet is a classical CNN-based network that performs well in variable computer vision tasks, especially for image classification[69]. We implement ResNet-50, whose architecture is strictly the same as in [69], but adopt a 3D volume as the input.

Vison transformer base (ViT-B): A ViT trained from scratch [17]. Similar to our encoder, it contains 12 transformer blocks and a classifier. Considering the large number of voxels in 3D volumes, we use 1000 as our hidden size instead of 768 in [17].

Masked autoencoder (MAE): A state-of-the-art self-supervised ViT-based model for ImageNet[21]. It has a ViT-encoder and a ViT-decoder, a projection layer is used to connect the encoder and decoder, and pixel loss is employed as the single loss function. To better compare the performance of our proposed model and the MAE, we use the same strategy and hyperparameters to train both the two models.

### 5) Experimental Settings

To verify the generalizability and robustness of our proposed CS-CRL, multiple classification tasks were utilized, such as AD classification (AD vs. NC), SZ classification (SZ vs. NC) and ADHD classification (ADHD vs. NC). Furthermore, we verify the transferability of our CS-CRL on three cross-dataset classification tasks. The classification performance was evaluated by four metrics, including the accuracy (ACC), sensitivity (SEN), specificity (SPE), and the area under the receiver operating characteristic curve (AUC).

TABLE III
WITHIN-DOMAIN CLASSIFICATION RESULTS

| Dataset | Method | ACC | SEN | SPE | AUC |
|---|---|---|---|---|---|
| ADNI | PLM | 0.8693 | 0.8088 | 0.9176 | 0.8632 |
| | DA-MIDL | 0.9085 | 0.8529 | 0.9529 | 0.9286 |
| | Resnet50 | 0.8758 | 0.7794 | 0.9529 | 0.9280 |
| | ViT | 0.8758 | 0.8382 | 0.9059 | 0.8989 |
| | MAE | 0.8693 | 0.7500 | **0.9647** | 0.9138 |
| | CS-CRL | **0.9216** | **0.8971** | 0.9412 | **0.9439** |
| MCIC | PLM | 0.6410 | 0.7619 | 0.5000 | 0.6310 |
| | DA-MIDL | 0.7436 | 0.6667 | 0.8333 | 0.8386 |
| | Resnet50 | 0.7436 | 0.7619 | 0.7222 | 0.7063 |
| | ViT | 0.7692 | 0.7619 | 0.7778 | 0.7354 |
| | MAE | 0.7949 | 0.7619 | 0.8333 | 0.7910 |
| | CS-CRL | **0.8718** | **0.8571** | **0.8889** | **0.8677** |
| NUSDAST | PLM | 0.6316 | **0.8947** | 0.3684 | 0.6316 |
| | DA-MIDL | 0.7368 | **0.8421** | 0.6316 | 0.7174 |
| | Resnet50 | 0.7368 | 0.7368 | 0.7368 | 0.6898 |
| | ViT | 0.6579 | **0.8421** | 0.4737 | 0.6953 |
| | MAE | 0.7895 | 0.7895 | 0.7895 | 0.7479 |
| | CS-CRL | **0.8158** | **0.8421** | 0.7895 | **0.7673** |
| NYU | PLM | 0.6889 | 0.8333 | 0.5238 | 0.6786 |
| | DA-MIDL | 0.6444 | 0.7917 | 0.4762 | 0.5675 |
| | Resnet50 | 0.7111 | **0.9583** | 0.4286 | 0.4901 |
| | ViT | 0.7333 | 0.6667 | **0.8095** | 0.7341 |
| | MAE | 0.7111 | **0.9583** | 0.4286 | 0.6806 |
| | CS-CRL | **0.7556** | 0.7083 | **0.8095** | **0.7579** |

TABLE IV
CLASSIFICATION RESULTS ON CROSS-DOMAIN TRANSFER LEARNING

| Source domain → Target domain | Method | ACC | SEN | SPE | AUC |
|---|---|---|---|---|---|
| ADNI → OASIS | PLM | 0.7631 | 0.6176 | 0.8912 | 0.7544 |
| | DA-MIDL | 0.8402 | 0.7941 | 0.8808 | 0.8682 |
| | Resnet50 | 0.8512 | 0.8059 | 0.8912 | 0.8857 |
| | ViT | 0.8760 | 0.8000 | 0.9430 | 0.9174 |
| | MAE | 0.8540 | **0.8588** | 0.8497 | 0.8997 |
| | CS-CRL | **0.8815** | 0.8059 | **0.9482** | **0.9181** |
| MCIC → NUSDAST | PLM | 0.5781 | 0.5833 | 0.5729 | 0.5798 |
| | DA-MIDL | 0.6302 | 0.6250 | 0.6354 | 0.6408 |
| | Resnet50 | 0.6510 | 0.5625 | **0.7396** | 0.6564 |
| | ViT | 0.6563 | 0.7396 | 0.5729 | 0.6663 |
| | MAE | 0.7135 | **0.8125** | 0.6146 | 0.7327 |
| | CS-CRL | **0.7344** | 0.7500 | 0.7188 | **0.7483** |
| NYU → KKI | PLM | 0.4727 | 0.4800 | 0.6000 | 0.4800 |
| | DA-MIDL | **0.6727** | 0.5200 | **0.8000** | 0.5898 |
| | Resnet50 | 0.6364 | 0.6800 | 0.6000 | 0.5593 |
| | ViT | 0.6545 | 0.7600 | 0.5667 | **0.6773** |
| | MAE | 0.6364 | 0.6800 | 0.6000 | 0.5227 |
| | CS-CRL | **0.6727** | **0.7600** | 0.6000 | 0.6649 |

## IV. RESULTS

### A. Within-domain Classification Performance

The same training, validation, and test set are used in the competing methods and our CS-CRL. The classification results are reported in Table 3. There are three observations. First, our proposed CS-CRL performed better on all four downstream classification tasks compared to the other methods. This result suggests that global contexts captured by CS-CRL are closely related to crucial intrinsic properties contributing to normal brain functional. Second, CS-CRL, ViT, ResNet-50 and DA-MIDL performed well in AD diagnosis. However, in more challenging tasks, such as the diagnosis of SZ and ADHD, our CS-CRL shows a prominent advantage over other methods. This result suggests that the inter-regional dependencies established by CS-CRL facilitates the detection of subtle lesions in interregional interactions that cannot be captured by the other methods. Third, when we repeat an overall survey for all four tasks, the performances of the ViT trained from scratch and the ViT pre-trained by MAE are not significantly better than those of the CNN-based ResNet-50 and DA-MIDL. This result further demonstrates that, without suitable proxy tasks and constraints, ViT-based models are hard to train due to its limited built-in architectural priors.

### B. Cross-domain Classification Performance

We evaluate the proposed CS-CRL and the competing methods in cross-domain tasks. Specifically, for AD classification (AD vs. NC), SZ classification (SZ vs. NC) and ADHD classification (ADHD vs. NC), we use one dataset as the source domain and another as the target domain to evaluate the transferability. The cross-domain classification results are reported in Table 4. From Table 4, one can observe that our proposed CS-CRL consistently outperforms the competing methods in cross-domain tasks, which demonstrates that our CS-CRL has better generalization, and is robust against the heterogeneity caused by different scanners, scanning protocols, and subject cohorts. In addition, limited by the scale of the datasets, the small test sets of the MCIC and NUSDAST datasets may affect the credibility of the classification result. The introduction of cross-domain task expands the size of the test set to 192 in SZ classification, and demonstrates the superiority of our CS-CRL in certainty.

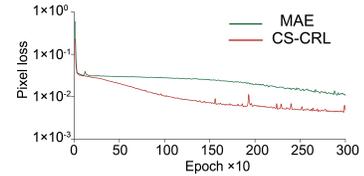

**Fig. 2.** Pixel loss during pre-training. Loss values are calculated using the mean square error (MSE).

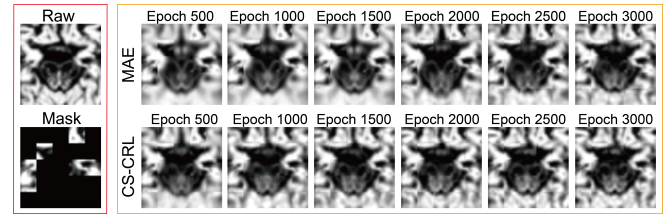

**Fig. 3.** Images reconstructed by CS-CRL and MAE in axial view. The first box shows the raw image and the random mask. The second box shows reconstructed images at different

training epochs. Overall, images from CS-CRL have better-delineated tissue boundaries and lower artifact/noise levels.

*C. Convergence and Reconstruction*

The pixel loss and visual comparison between the original and reconstructed images in the pre-training stages of the MAE and CS-CRL are shown in Figs. 2 and 3. It is noticeable that our proposed CS-CRL converges faster and more precisely than MAE. Images reconstructed by CS-CRL have better-delineated tissue boundaries and lower artifact and noise levels than those reconstructed by the MAE. The image quality of the CS-CRL trained for 1000 epochs equaled or surpassed those of the MAE at 3000 epochs. This finding suggests that latent representation of CS-CRL better aggregates the global context while taking into account the details, from which the decoder can easily deduce information in masked patches.

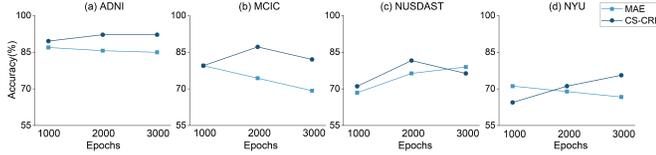

**Fig. 4.** Within-domain classification performance of MAE and CS-CRL. (a) - (d) present the classification accuracy on the four datasets.

*D. Effectiveness of Epochs in the Pre-training Stage*

To explore how pre-training epochs affect the representation learning, we analyzed the performance of CS-CRL models pre-trained for 1000, 2000, and 3000 epochs on downstream tasks, using the MAE as the baseline (Fig. 4). Overall, our CS-CRL performed better than MAE at almost any pre-training epoch, which further suggests the effectiveness and stability of our CS-CRL. The CS-CRL model pre-trained for 2000 epochs achieved the best performance on AD and SZ classification tasks. Our CS-CRL training time is approximately 125 epochs per hour, using 4 GPUs. As a result, it takes only approximately 16 hours to pre-train a high-performance and high-generalization model.

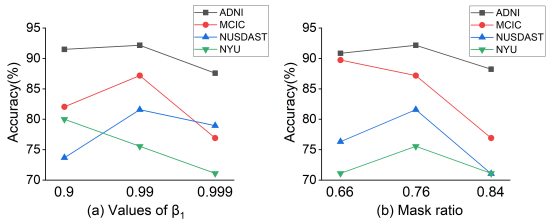

**Fig. 5.** Within-domain classification results on different parameter settings. (a) $\beta_1$ settings (b) Mask ratio settings.

*E. Effectiveness of Parameters in the Pre-training Stage*

The parameters $\beta_1$ and $\beta_2$ (section III.A.(6)) and mask ratio (section III.A.(2)) have an obvious influence on the convergence and performance of our model. In Fig. 5, we vary the value of $\beta_1$ in the range of (0.9, 0.99, 0.999) and the value of the mask ratio in the range of (0.66, 0.76, 0.84) to study the effectiveness of the parameters. Note that the value of $\beta_2$ is always kept at $1/2(1-\beta_1)$. From Fig. 5, we observe that our

CS-CRL can achieve better performance with $\beta_1 \in (0.9, 0.99)$ and a mask ratio of 0.76, which is consistent with most of MIM methods in which the mask ratio is suggested to be chosen within the range of (0.70,0.80) [21]. Additionally, with $\beta_1 = 0.999$, the performance of CS-CRL is not that good. This suggests that the biological prior plays an important role in the pretraining stage, and weakening this regularization, which is specific to brain sMRI, will impair the performance of CS-CRL to a great extent.

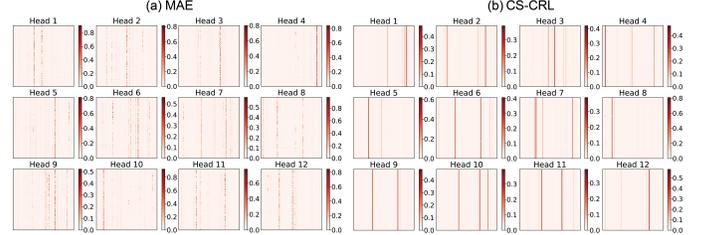

**Fig. 6.** (a) Attention map of layer 1 in the MAE pre-trained for 1000 epochs. (b) Attention map of layer 1 in our CS-CRL pre-trained for 2000 epochs. In all matrices, the X-axis indicates the target region, and the Y-axis indicates the source region. Each edge indicates the dependence of the source region on the target region. The attention maps are: $softmax(QK^T/\sqrt{d_k})$.

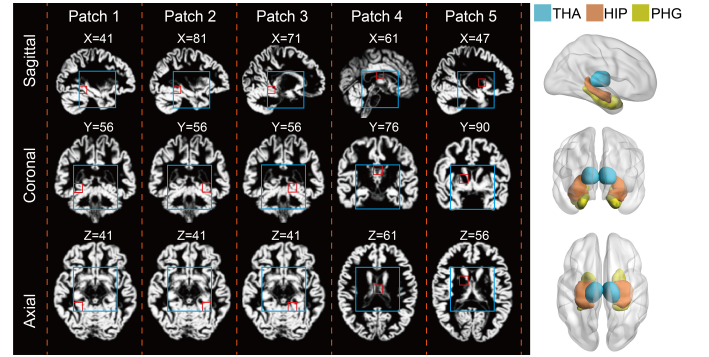

**Fig. 7.** Locations of hub patches in the brain. Red boxes indicate hub patches and blue boxes indicate the subcortical structure used in our study. The last column shows the atlas of the regions that hub patches mainly cover. THA: Thalamus, HIP: Hippocampus, PHG: Parahippocampal gyrus.

## V. DISCUSSION

*A. Visualization of Attention Maps and Patch Locations*

We attempt to reveal the internal schema for the aggregation of global context in the encoder mentioned above by extracting the attention maps from MSA blocks in the same layer during a forward propagation process. Considering the performance of pre-trained models, we select the model pre-trained for 2000 epochs by CS-CRL for visualization and one pre-trained for 1000 epochs by MAE for comparison. The attention maps are averaged across subjects in IXI dataset (Fig. 6). For CS-CRL, the distribution of attention maps is regular, which means that patches pay significant attention to the global context. For MAE, the distribution of the attention maps is slightly messy, indicating that fewer patches focus on the global context and more patches focus on local information. This suggests that there are a number of highly connected hub regions in the brain, and our CS-CRL captured this connectional property and establish credible dependency between hubs and other regions.

We calculate the averaged attention map across heads and then determine the most critical patches according to the sum value of each column. The value of one column is positively related to the dependency between the given patch and others. We chose the 5 most important patches and visualized their locations in the brain (Fig. 7). These patches mainly covered the hippocampus and parahippocampal gyrus and a small portion of the thalamus. These regions are closely related to cognition and memory and brain disorders [70, 71]. Thus, our model automatically locates key brain regions and finds that the remaining regions are strongly dependent on them. Actually, the good performance of our CS-CRL is biologically explicable.

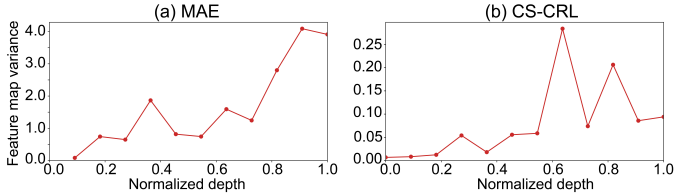

**Fig. 8.** Variance of feature maps from the 12 transformer blocks of the MAE and CS-CRL. For the possible comparison between models of different depths in the future, we choose the normalized depth to plot the results. Note that the variance of the first block in the MAE is too large to be shown in this figure.

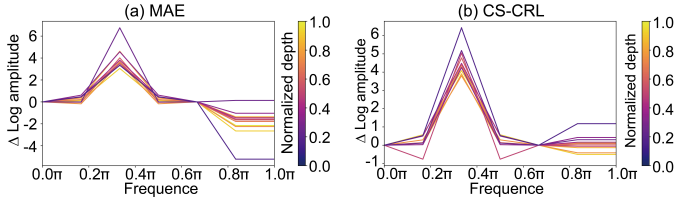

**Fig. 9.** Relative log amplitudes of the Fourier transformed feature maps. Δ Log amplitude of high-frequency signals is the difference between the log amplitude at normalized frequency $0.0\pi$ (center) and at $1.0\pi$ (boundary). For better visualization, we only provide the first 7 diagonal components of the Fourier transformed feature maps.

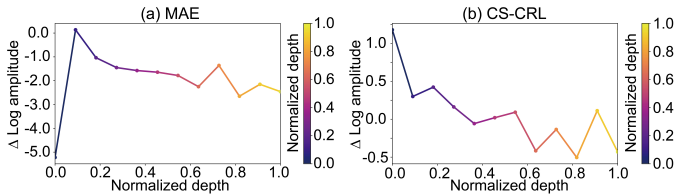

**Fig. 10.** The Δ log amplitude of the MAE and CS-CRL at high-frequency ($1.0\pi$), showing that CS-CRL tends to reduce the high-frequency signal, but the MAE does not.

### B. Variance and Fourier Analysis of Feature Maps

We are motivated by an assumption that the connectional property is a general characteristic of the brain, and multiple brain diseases, whether they cause obvious structural changes or not, lead to connectional abnormalities in the brain. In theory, using the connectional prior as regularization can guide the model to extract similar semantic representations from all of the samples. Specifically, for different samples, their feature maps should be distributed in a stable and compressed latent space, and similar frequency components of their Fourier transformed feature maps should be preferred or declined by the model. The following results showed a reverse verification for this assumption in practice.

We selected the same models as in Fig. 6 and extracted feature maps from the 12 transformer blocks. The feature maps are averaged on a minibatch of samples and visualized in Fig. 8. The fact is that the variances of the feature maps in CS-CRL are much smaller than those in MAE, suggesting that CS-CRL better aggregates features and establishes the latent space more precisely.

The Fourier analysis of the feature maps (Fig. 9) shows that CS-CRL tends to reduce high-frequency signals layer by layer, while the MAE does not show a strong preference for low-frequency signals or high-frequency signals. See Fig. 10 for a detailed analysis. It suggests that the performance of our CS-CRL partly resulted from the following two factors: robustness against high-frequency noise and accurate extraction of the specific frequency components in sMRI, which can be seen as general biomarkers to diagnose multiple brain diseases.

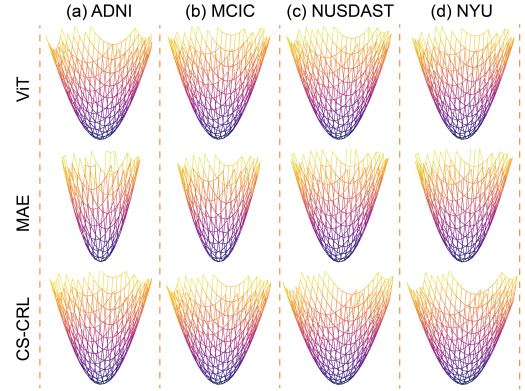

**Fig. 11.** Loss landscape visualizations. We calculate the loss (NLL + L2 regularization) by looping over all the data points in the training sets in four datasets using models that perform best during evaluation.

### C. Visualization of Loss Landscapes

In section V.A, we demonstrated that our proposed CS-CRL could capture the general characteristics across diseases and datasets from the perspective of feature analysis. Here, we analyzed the training loss and further confirmed that benefiting from the biological prior, our CS-CRL learned general representations of the training samples, and thus had better trainability and was easier to optimize.

In Fig. 11, We used filter normalization [72] to visualize the loss landscapes. Previous studies have demonstrated that the flatter the loss landscape is, the better the generalization and robustness[63]. Therefore, the loss landscape is a reliable indicator for comparison across models. We reported a thought-provoking finding that MAE has a sharpen loss landscape than ViT, while CS-CRL has a much flatten one. In the diagnosis of SZ-S, the model pre-trained by MAE gained a much higher accuracy than the ViT trained from scratch, but its loss landscape is still sharper, indicating that the loss landscapes represent ease of optimization and are irrelevant to the accuracy of classification, as mentioned in [72].

Actually, the MAE minimizes the size of the connected region around the minimum where the training loss remains low, leading to difficulty in stable training. In contrast, our CS-CRL has flatter loss landscapes, indicating that it not only achieves superior performance on classification tasks but also shows advantages in generalization and robustness. Therefore, CS-CRL demonstrated its ease of training and insensitivity to hyperparameters over others ViT-based models.

### D. Advantages of CS-CRL

It is widely accepted that ViTs are apt at capturing global context benefit from their weak inductive bias [73]. However, the weak inductive bias and lack of built-in architecture priors disrupt ViT training [17]. In this study, the ViT trained from scratch did not achieve satisfactory performance, suggesting that the long-term dependence it learned did not match the connectional pattern in the brain. Our proposed CS-CRL addresses this problem by integrating computational and biological principles into a single framework. We showed that CS-CRL leveraged only one healthy people dataset for pre-training and achieved superior performance on multiple patient datasets. The results suggest that the connectional regularization combined with the mask reconstruction task enables the ViT encoder to learn multilevel and nonredundant information and to effectively establish inter-regional dependencies with biological plausibility. Thus, CS-CRL essentially learns a biological inductive bias that applies to most brain sMRI data domains. The CS-CRL is conceptually similar to VAE [74]. VAE uses two projection layers to fit the 'mean' and 'variance' and then samples the potential representation, which conforms to the normal distribution, while our CS-CRL uses Gram matrices to constrain the latent representations. We are inspired by the same idea: constructing a prior assumption to regularize the model. Besides, we revealed an instructive conclusion from theory and practice: in brain medical imaging, self-supervised pre-training with additional proper biological priors can significantly improve understanding of the intrinsic properties in the brain, which may be more important and meaningful than modifying the architecture of the backbone.

### E. Prospect of CS-CRL

In AD diagnosis task, our CS-CRL outperforms the state-of-the-art CNN-based models. Local atrophy in the AD brain is obvious and is an important diagnostic basis, while our study shows that global changes caused by local atrophy may be more important and discriminative, which is a key breakthrough for the diagnosis of AD. Therefore, models based on our CS-CRL are promising in AD and its prodromal stage data analysis.

In addition, the local abnormalities of SZ and ADHD are not obvious, leading to dependency on fMRI features of most CAD models. However, fMRI-based models require complex data preprocessing, feature extraction and feature determination by statistical analysis. These steps add implementation complexity and time cost. Moreover, different preprocessing methods have great influences on the performance of CAD models, causing poor generalization and robustness. Even worse, selecting features by statistical analysis may lead to information leakage. Our proposed CS-CRL performed well on SZ and ADHD diagnostic tasks using sMRI data, which is competitive with advanced fMRI-based CAD methods [75-77], providing a more effective method for the diagnosis of SZ and ADHD.

There are several inspirations that may help further improve CS-CRL performance. First, only one patch size was used and we can further explore the influence of patch size to determine the optimal one. Second, ViT-based models are memory consuming and computationally intensive. Therefore, we can design a pyramid or hierarchical architecture for a lightweight model. Third, given the importance of biological priors, it is a direct idea to design a data enhancement method that is specific to 3D sMRI data and could be in conjunction with our CS-CRL.

## VI. CONCLUSION

In this study, we propose a representation learning model for brain sMRI, CS-CRL, which does not need any extra domain-adaptation module but shows advantages in accuracy, generalization and explanation. Our CS-CRL learns strong representations with biological significance that capture connectional patterns in the brain. These representations are robust across data domains and is applicable to various brain disease diagnostic tasks.

## REFERENCES

[1]  X. Liu, *et al.*, "Self-supervised learning: Generative or contrastive," *IEEE Transactions on Knowledge and Data Engineering,* 2021.
[2]  S. d'Ascoli, H. Touvron, M. L. Leavitt, A. S. Morcos, G. Biroli, and L. Sagun, "Convit: Improving vision transformers with soft convolutional inductive biases," in *International Conference on Machine Learning*, 2021, pp. 2286-2296.
[3]  N. Park and S. Kim, "How Do Vision Transformers Work?," in *International Conference on Learning Representations*, 2021.
[4]  S. V. M. Sagheer and S. N. George, "A review on medical image denoising algorithms," *Biomedical signal processing and control,* vol. 61, p. 102036, 2020.
[5]  R. Geirhos, P. Rubisch, C. Michaelis, M. Bethge, F. A. Wichmann, and W. Brendel, "ImageNet-trained CNNs are biased towards texture; increasing shape bias improves accuracy and robustness," in *International Conference on Learning Representations*, 2018.
[6]  C. Lian, M. Liu, J. Zhang, and D. Shen, "Hierarchical fully convolutional network for joint atrophy localization and Alzheimer's disease diagnosis using structural MRI," *IEEE transactions on pattern analysis and machine intelligence,* vol. 42, pp. 880-893, 2018.
[7]  W. Zhu, L. Sun, J. Huang, L. Han, and D. Zhang, "Dual attention multi-instance deep learning for Alzheimer's disease diagnosis with structural MRI," *IEEE Transactions on Medical Imaging,* vol. 40, pp. 2354-2366, 2021.
[8]  M. Liu, J. Zhang, E. Adeli, and D. Shen, "Landmark-based deep multi-instance learning for brain disease diagnosis," *Medical image analysis,* vol. 43, pp. 157-168, 2018.
[9]  H. Guan, Y. Liu, E. Yang, P.-T. Yap, D. Shen, and M. Liu, "Multi-site MRI harmonization via attention-guided deep domain adaptation for brain disorder identification," *Medical Image Analysis,* vol. 71, p. 102076, 2021.
[10] D. Lei, *et al.*, "Integrating machining learning and multimodal neuroimaging to detect schizophrenia at the level of the individual," *Human Brain Mapping,* vol. 41, pp. 1119-1135, 2020.
[11] M. D. Rosenberg and E. S. Finn, "How to establish robust brain–behavior relationships without thousands of individuals," *Nature Neuroscience,* vol. 25, pp. 835-837, 2022.
[12] E. Bullmore and O. Sporns, "The economy of brain network organization," *Nature reviews neuroscience,* vol. 13, pp. 336-349, 2012.
[13] B. M. Tijms, P. Seriès, D. J. Willshaw, and S. M. Lawrie, "Similarity-based extraction of individual networks from gray matter MRI scans," *Cerebral cortex,* vol. 22, pp. 1530-1541, 2012.

[14] J. Seidlitz, et al., "Morphometric similarity networks detect microscale cortical organization and predict inter-individual cognitive variation," *Neuron,* vol. 97, pp. 231-247. e7, 2018.

[15] Y. Jiang, et al., "Progressive reduction in gray matter in patients with schizophrenia assessed with MR imaging by using causal network analysis," *Radiology,* vol. 287, pp. 633-642, 2018.

[16] W. Zhang, et al., "Brain gray matter network organization in psychotic disorders," *Neuropsychopharmacology,* vol. 45, pp. 666-674, 2020.

[17] A. Dosovitskiy, et al., "An Image is Worth 16x16 Words: Transformers for Image Recognition at Scale," in *International Conference on Learning Representations*, 2020.

[18] A. Vaswani, et al., "Attention is all you need," *Advances in neural information processing systems,* vol. 30, 2017.

[19] M. M. Naseer, K. Ranasinghe, S. H. Khan, M. Hayat, F. Shahbaz Khan, and M.-H. Yang, "Intriguing properties of vision transformers," *Advances in Neural Information Processing Systems,* vol. 34, pp. 23296-23308, 2021.

[20] R. Shao, Z. Shi, J. Yi, P.-Y. Chen, and C.-J. Hsieh, "On the adversarial robustness of vision transformers," *arXiv preprint arXiv:2103.15670,* 2021.

[21] K. He, X. Chen, S. Xie, Y. Li, P. Dollár, and R. Girshick, "Masked autoencoders are scalable vision learners," *arXiv preprint arXiv:2111.06377,* 2021.

[22] H. Bao, L. Dong, and F. Wei, "Beit: Bert pre-training of image transformers," *arXiv preprint arXiv:2106.08254,* 2021.

[23] X. Dong, et al., "PeCo: Perceptual Codebook for BERT Pre-training of Vision Transformers," *arXiv preprint arXiv:2111.12710,* 2021.

[24] L. Liu, et al., "Cortical abnormalities and identification for first-episode schizophrenia via high-resolution magnetic resonance imaging," *Biomarkers in Neuropsychiatry,* vol. 3, p. 100022, 2020.

[25] A. Talpalaru, N. Bhagwat, G. A. Devenyi, M. Lepage, and M. M. Chakravarty, "Identifying schizophrenia subgroups using clustering and supervised learning," *Schizophrenia research,* vol. 214, pp. 51-59, 2019.

[26] W. Shao, Y. Peng, C. Zu, M. Wang, D. Zhang, and A. s. D. N. Initiative, "Hypergraph based multi-task feature selection for multimodal classification of Alzheimer's disease," *Computerized Medical Imaging and Graphics,* vol. 80, p. 101663, 2020.

[27] R. Iannaccone, T. U. Hauser, J. Ball, D. Brandeis, S. Walitza, and S. Brem, "Classifying adolescent attention-deficit/hyperactivity disorder (ADHD) based on functional and structural imaging," *European child & adolescent psychiatry,* vol. 24, pp. 1279-1289, 2015.

[28] M. Liu, D. Zhang, and D. Shen, "Relationship induced multi-template learning for diagnosis of Alzheimer's disease and mild cognitive impairment," *IEEE transactions on medical imaging,* vol. 35, pp. 1463-1474, 2016.

[29] Y. Pan, et al., "Morphological profiling of schizophrenia: cluster analysis of MRI-based cortical thickness data," *Schizophrenia bulletin,* vol. 46, pp. 623-632, 2020.

[30] T. Jo, K. Nho, and A. J. Saykin, "Deep learning in Alzheimer's disease: diagnostic classification and prognostic prediction using neuroimaging data," *Frontiers in aging neuroscience,* vol. 11, p. 220, 2019.

[31] D. Sadeghi, et al., "An overview of artificial intelligence techniques for diagnosis of Schizophrenia based on magnetic resonance imaging modalities: Methods, challenges, and future works," *Computers in Biology and Medicine,* p. 105554, 2022.

[32] A. S. Heinsfeld, A. R. Franco, R. C. Craddock, A. Buchweitz, and F. Meneguzzi, "Identification of autism spectrum disorder using deep learning and the ABIDE dataset," *NeuroImage: Clinical,* vol. 17, pp. 16-23, 2018.

[33] C. Lian, M. Liu, Y. Pan, and D. Shen, "Attention-guided hybrid network for dementia diagnosis with structural MR images," *IEEE transactions on cybernetics,* 2020.

[34] M. Hu, K. Sim, J. H. Zhou, X. Jiang, and C. Guan, "Brain MRI-based 3D convolutional neural networks for classification of schizophrenia and controls," in *2020 42nd Annual International Conference of the IEEE Engineering in Medicine & Biology Society (EMBC)*, 2020, pp. 1742-1745.

[35] Z. Wang, Y. Sun, Q. Shen, and L. Cao, "Dilated 3D convolutional neural networks for brain MRI data classification," *IEEE Access,* vol. 7, pp. 134388-134398, 2019.

[36] J. Li, J. Chen, Y. Tang, B. A. Landman, and S. K. Zhou, "Transforming medical imaging with Transformers? A comparative review of key properties, current progresses, and future perspectives," *arXiv preprint arXiv:2206.01136,* 2022.

[37] J. Chen, et al., "Transunet: Transformers make strong encoders for medical image segmentation," *arXiv preprint arXiv:2102.04306,* 2021.

[38] G. Xu, X. Wu, X. Zhang, and X. He, "Levit-unet: Make faster encoders with transformer for medical image segmentation," *arXiv preprint arXiv:2107.08623,* 2021.

[39] H. Cao, et al., "Swin-unet: Unet-like pure transformer for medical image segmentation," *arXiv preprint arXiv:2105.05537,* 2021.

[40] Z. Liu, et al., "Swin transformer: Hierarchical vision transformer using shifted windows," in *Proceedings of the IEEE/CVF International Conference on Computer Vision*, 2021, pp. 10012-10022.

[41] H.-C. Shin, et al., "Ganbert: Generative adversarial networks with bidirectional encoder representations from transformers for mri to pet synthesis," *arXiv preprint arXiv:2008.04393,* 2020.

[42] J. Devlin, M.-W. Chang, K. Lee, and K. Toutanova, "Bert: Pre-training of deep bidirectional transformers for language understanding," *arXiv preprint arXiv:1810.04805,* 2018.

[43] S. A. Kamran, K. F. Hossain, A. Tavakkoli, S. L. Zuckerbrod, and S. A. Baker, "Vtgan: Semi-supervised retinal image synthesis and disease prediction using vision transformers," in *Proceedings of the IEEE/CVF International Conference on Computer Vision*, 2021, pp. 3235-3245.

[44] B. Gheflati and H. Rivaz, "Vision transformer for classification of breast ultrasound images," *arXiv preprint arXiv:2110.14731,* 2021.

[45] K. S. Krishnan and K. S. Krishnan, "Vision transformer based COVID-19 detection using chest X-rays," in *2021 6th International Conference on Signal Processing, Computing and Control (ISPCC)*, 2021, pp. 644-648.

[46] Y. Dai, Y. Gao, and F. Liu, "Transmed: Transformers advance multi-modal medical image classification," *Diagnostics,* vol. 11, p. 1384, 2021.

[47] X. Gao, Y. Qian, and A. Gao, "Covid-vit: Classification of covid-19 from ct chest images based on vision transformer models," *arXiv preprint arXiv:2107.01682,* 2021.

[48] H.-C. Shin, et al., "Deep convolutional neural networks for computer-aided detection: CNN architectures, dataset characteristics and transfer learning," *IEEE transactions on medical imaging,* vol. 35, pp. 1285-1298, 2016.

[49] N. Tajbakhsh, et al., "Convolutional neural networks for medical image analysis: Full training or fine tuning?," *IEEE transactions on medical imaging,* vol. 35, pp. 1299-1312, 2016.

[50] A. K. Mondal, A. Bhattacharjee, P. Singla, and A. Prathosh, "xViTCOS: explainable vision transformer based COVID-19 screening using radiography," *IEEE Journal of Translational Engineering in Health and Medicine,* vol. 10, pp. 1-10, 2021.

[51] H. Spitzer, K. Kiwitz, K. Amunts, S. Harmeling, and T. Dickscheid, "Improving cytoarchitectonic segmentation of human brain areas with self-supervised siamese networks," in *International Conference on Medical Image Computing and Computer-Assisted Intervention*, 2018, pp. 663-671.

[52] C. Matsoukas, J. F. Haslum, M. Söderberg, and K. Smith, "Is it time to replace cnns with transformers for medical images?," *arXiv preprint arXiv:2108.09038,* 2021.

[53] M. Caron, et al., "Emerging properties in self-supervised vision transformers," in *Proceedings of the IEEE/CVF International Conference on Computer Vision*, 2021, pp. 9650-9660.

[54] A. Taleb, et al., "3d self-supervised methods for medical imaging," *Advances in Neural Information Processing Systems,* vol. 33, pp. 18158-18172, 2020.

[55] Z. Zhou, et al., "Models genesis: Generic autodidactic models for 3d medical image analysis," in *International conference on medical image computing and computer-assisted intervention*, 2019, pp. 384-393.

[56] L. A. Gatys, A. S. Ecker, and M. Bethge, "Image style transfer using convolutional neural networks," in *Proceedings of the IEEE conference on computer vision and pattern recognition*, 2016, pp. 2414-2423.

[57] L. Gatys, A. S. Ecker, and M. Bethge, "Texture synthesis using convolutional neural networks," *Advances in neural information processing systems,* vol. 28, 2015.

[58] T. N. Kipf and M. Welling, "Variational graph auto-encoders," *arXiv preprint arXiv:1611.07308,* 2016.

[59] C. R. Jack Jr, et al., "The Alzheimer's disease neuroimaging initiative (ADNI): MRI methods," *Journal of Magnetic Resonance Imaging: An Official Journal of the International Society for Magnetic Resonance in Medicine,* vol. 27, pp. 685-691, 2008.

[60] P. J. LaMontagne, et al., "OASIS-3: longitudinal neuroimaging, clinical, and cognitive dataset for normal aging and Alzheimer disease," *MedRxiv,* 2019.

[61] R. L. Gollub, *et al.*, "The MCIC collection: a shared repository of multi-modal, multi-site brain image data from a clinical investigation of schizophrenia," *Neuroinformatics,* vol. 11, pp. 367-388, 2013.

[62] L. Wang, *et al.*, "Northwestern University schizophrenia data and software tool (NUSDAST)," *Frontiers in neuroinformatics,* vol. 7, p. 25, 2013.

[63] A.-. consortium, "The ADHD-200 consortium: a model to advance the translational potential of neuroimaging in clinical neuroscience," *Frontiers in systems neuroscience,* vol. 6, p. 62, 2012.

[64] J. Ashburner and K. J. Friston, "Voxel-based morphometry—the methods," *Neuroimage,* vol. 11, pp. 805-821, 2000.

[65] M. Jenkinson, C. F. Beckmann, T. E. Behrens, M. W. Woolrich, and S. M. Smith, "Fsl," *Neuroimage,* vol. 62, pp. 782-790, 2012.

[66] B. U. Forstmann, G. de Hollander, L. van Maanen, A. Alkemade, and M. C. Keuken, "Towards a mechanistic understanding of the human subcortex," *Nature Reviews Neuroscience,* vol. 18, pp. 57-65, 2017.

[67] O. Sporns, "Network attributes for segregation and integration in the human brain," *Current opinion in neurobiology,* vol. 23, pp. 162-171, 2013.

[68] P. T. Bell and J. M. Shine, "Subcortical contributions to large-scale network communication," *Neuroscience & Biobehavioral Reviews,* vol. 71, pp. 313-322, 2016.

[69] K. He, X. Zhang, S. Ren, and J. Sun, "Deep residual learning for image recognition," in *Proceedings of the IEEE conference on computer vision and pattern recognition*, 2016, pp. 770-778.

[70] L. W. de Jong, *et al.*, "Strongly reduced volumes of putamen and thalamus in Alzheimer's disease: an MRI study," *Brain,* vol. 131, pp. 3277-3285, 2008.

[71] J. P. Aggleton, S. M. O'Mara, S. D. Vann, N. F. Wright, M. Tsanov, and J. T. Erichsen, "Hippocampal–anterior thalamic pathways for memory: uncovering a network of direct and indirect actions," *European Journal of Neuroscience,* vol. 31, pp. 2292-2307, 2010.

[72] H. Li, Z. Xu, G. Taylor, C. Studer, and T. Goldstein, "Visualizing the loss landscape of neural nets," *Advances in neural information processing systems,* vol. 31, 2018.

[73] J.-B. Cordonnier, A. Loukas, and M. Jaggi, "On the Relationship between Self-Attention and Convolutional Layers," in *Eighth International Conference on Learning Representations-ICLR 2020*, 2020.

[74] D. P. Kingma and M. Welling, "Auto-encoding variational bayes," *arXiv preprint arXiv:1312.6114,* 2013.

[75] W. Yan, *et al.*, "Discriminating schizophrenia using recurrent neural network applied on time courses of multi-site FMRI data," *EBioMedicine,* vol. 47, pp. 543-552, 2019.

[76] L.-L. Zeng, *et al.*, "Multi-site diagnostic classification of schizophrenia using discriminant deep learning with functional connectivity MRI," *EBioMedicine,* vol. 30, pp. 74-85, 2018.

[77] T. Wang, A. Bezerianos, A. Cichocki, and J. Li, "Multikernel capsule network for schizophrenia identification," *IEEE transactions on Cybernetics,* 2020.